\documentclass{ws-p8-50x6-00}

\newcommand{\Ham}{{\cal H}}
\newcommand{\vecr}{{\mathbf r}}

\def\ARPC{\em Annu. Rev. Phys. Chem.}
\def\PRA{{\em Phys. Rev.} A}
\def\PRB{{\em Phys. Rev.} B}
\def\JPSJ{\em J. Phys. Soc. Japan}

\usepackage{amsmath,graphicx,bm}

\begin{document}

\title{Atomic spectra in a helium bubble}

\author{Takashi Nakatsukasa}

\address{Physics Department, Tohoku University, Sendai 980-8578, Japan
}

\author{Kazuhiro Yabana}

\address{Institute of Physics,
University of Tsukuba, Tsukuba 305-8571, Japan
}

\author{George F. Bertsch}

\address{Physics Department and Institute for Nuclear Theory,
University of Washington, Seattle, WA 98195, USA
}


\maketitle

\abstracts{
Density functional theory (DFT) is applied to
atomic spectra under perturbations of superfluid liquid helium.
The atomic DFT of helium is used to obtain the distribution
of helium atoms around the impurity atom, and the electronic DFT is
applied to the excitations of the atom,
averaging over the ensemble of helium configurations.
The shift and broadening of the
$D_1$ and $D_2$ absorption lines are quite well reproduced by theory,
suggesting that the DFT may be useful for describing
spectral perturbations in more complex environments.
}

\section{Spectroscopy in liquid helium}
\label{sec:spectroscopy}

Spectroscopic measurements of impurity atoms and
molecules in superfluid helium have been attracting considerable
interest in recent years.\cite{TV98,Cal01}
The repulsive force between an impurity and
helium atoms induces a bubble around the impurity.
This leads to
a weak perturbation of helium atoms on the spectra of impurities.
The line shifts and spectral shapes induced by the helium perturbation
provide information on the properties of the bubble in the quantum liquid
as well as the excited states of the impurity.
Since the perturbation is weak,
this method also provides a unique tool for spectroscopic
measurements
of atomic clusters at low temperature.\cite{Cal01}

Because of its simplicity, perturbations on alkali-atom lines
have been studied extensively.\cite{TV98,KFTY95,KFY96}
For cesium (Cs) atoms,
there are two $s$-to-$p$
transitions, the
$D_1$ ($s_{1/2}\rightarrow p_{1/2}$)
and $D_2$ ($s_{1/2}\rightarrow p_{3/2}$) lines, both of
which are blue-shifted and acquire widths in a helium bath.
The shifts and widths of the two lines are different, and
the $D_2$ line
has a skewed shape suggesting a double-peak structure. These features
were first analyzed with
a collective vibration model of the helium bubble.\cite{KFTY95,KFY96}
That model reproduced average peak shifts, but gave
line widths less than a half of observed ones.
A more sophisticated analysis has been made treating the
liquid helium environment by the Path-Integral Monte-Carlo method.\cite{Oga99}
However, the method is very costly in computer resources and
is difficult to apply to more complex systems.
We will show that a density functional theory (DFT) together with a
statistical treatment of helium configurations
provides a simple and quantitative description for the helium perturbations.

\section{Application of DFT to impurity spectroscopy}

\subsection{DFT-plus-statistical description of liquid helium}

The energy of liquid helium in the DFT is assumed to
have the form,
$E=\int d\vecr \Ham_0(\vecr)$,
where we adopt the Orsay-Paris functional,\cite{Dup90}
\begin{equation}
\label{H_0}
\Ham_0(\vecr) = \frac{1}{2m}\left| \nabla\sqrt{\rho(\vecr)} \right|^2
 + \frac{1}{2}\int d\vecr' \rho(\vecr)\rho(\vecr') V_{\rm LJ}(|\vecr-\vecr'|)
 +\frac{c}{2}\rho(\vecr)\left(\bar\rho_\vecr\right)^{1+\gamma} .
\end{equation}
Here, $m$ is the mass of a helium atom and $\bar\rho_\vecr$ is
a coarse-grained density, and $V_{\rm LJ}$ is a screened
Lennard-Jones potential.

The effect of the impurity was treated by
including in Eq. (\ref{H_0}) a potential interaction, $V_{\rm I}(r)$,
between the helium atoms and the impurity,
\begin{equation}
\Ham(\vecr) = \Ham_0(\vecr) + V_{\rm I}(\vecr)\rho(\vecr).
\end{equation}
We approximate the $V_{\rm I}(\vecr)$ as a contact interaction,
\begin{equation}
V_{\rm I}(\vecr)
 = \frac{2\pi a}{m_{\rm e}} \rho_{\rm e}(\vecr) ,
\end{equation}
where $m_e$ is the electron mass and $\rho_{\rm e}(\vecr)$
is the electron density of the impurity which is calculated with
the electronic DFT in Sec.~\ref{subsec: helium_perturbation}.
The scattering length, $a$, 
is determined from the observed low-energy electron-helium
cross section.

Utilizing the energy functional, $E[\rho]=\int d\vecr \Ham(\vecr)$,
we calculate the density profile of liquid helium, putting the impurity
atom at the origin.
Minimizing the grand potential at zero temperature,
$\Omega\equiv E[\rho(\vecr)] - \mu N$, leads to a Hartree-type
equation
\begin{equation}
\label{Hartree}
\left[ -\frac{1}{2m} \nabla^2 + U(\vecr) + V_{\rm I}(\vecr) \right]
  \sqrt{\rho(\vecr)} = \mu \sqrt{\rho(\vecr)} .
\end{equation}
The equation is solved with the boundary condition that the density
go to the bulk density $\rho_0$ at large $r$.
Results indicate a sharp rise in the helium density at $r\approx 6$ \AA.
This corresponds to the bubble radius.

We use the $\rho(\vecr)$ computed above to generate an ensemble of
configurations of helium atoms as follows.
Take a large volume surrounding the alkali atom and
denote it as $V$. This volume includes $N$ helium atoms on average,
where $N$ is given by $\int_V d\vecr \rho(\vecr) = N$.
We randomly sample $N$ helium positions in $V$ according to the density
distribution $\rho(\vecr)$.
This sampling procedure gives probability distribution,
$w(\vecr_1,\cdots,\vecr_N)=\Pi_{i=1}^N (\rho(\vecr_i)/N)$.

\subsection{Helium perturbation on spectra}
\label{subsec: helium_perturbation}

Orbital wave functions of valence electrons in impurity, $\psi(\vecr)$,
are calculated using DFT
with Dirac wave functions and kinetic energy operator.
We need accurate wave functions at large distances from the atom,
which cannot be achieved with the traditional LDA functional due
to the incorrect orbital eigenvalues and the incorrect asymptotic
behavior of the potential. 
These problems are diminished with the gradient correction
which was designed to produce the correct asymptotic behaviour
of the potential.

We use first-order perturbation theory to evaluate the orbital shifts
in the ensemble of helium configurations
$\bm{\tau}=(\vecr_1,\cdots,\vecr_N)$.
The same helium configuration is used for the ground state $s_{1/2}$
and excited states $p_{1/2}$ and $p_{3/2}$, following the Frank-Condon
principle. For $s_{1/2}$ and $p_{1/2}$ states, the energy shifts of the
valence electron is then calculated as
$ \Delta E^{(k)}(\bm{\tau}) = 
(2\pi a/m_{\rm e})\sum_i |\psi^{(k)}(\vecr_i)|^2$,
where $k$ stands for orbital quantum numbers $(\ell j)$ and either
$m$ state may be taken.  For $p_{3/2}$ states, the matrix elements
depend on $m$ and we have to diagonalize a $4 \times 4$ matrix to get
the energy shifts.  We then obtain two eigenenergies, each of which is doubly
degenerate.

Each helium configuration produces an energy shift and possible splitting
but the transitions remain sharp.  The line broadening
comes from the ensemble average over helium configurations.
The line shape of the $D_1$ ($s_{1/2} \rightarrow p_{1/2}$)
transition is given by
\begin{equation}
S_{D_1}(E) = \int_V d\bm{\tau}
w(\bm{\tau})
\delta \left( E - \left( \Delta E^{(p_{1/2})}(\bm{\tau})
- \Delta E^{(s_{1/2})}(\bm{\tau}) \right) \right) ,
\label{defspectrum_1}
\end{equation}
where $E$ is a shift from the energy position of the free atom.
For the $D_2$ ($s_{1/2} \rightarrow p_{3/2}$) transition,
we have a similar expression but need to add the two eigenmodes.
\begin{figure}[ht]
\centerline{
\includegraphics[width=0.75\textwidth]{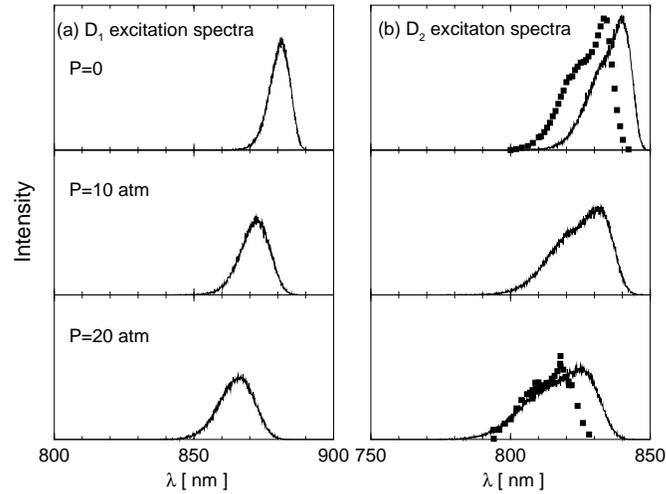}
}
\caption{\label{fig: No_correlation}
(a) Cs $D_1$ excitation spectrum at different helium pressure;
$P=0$, 10, and 20 atm.
(b) The same as (a) but for $D_2$ excitation spectrum.  Experimental
data\protect\cite{KFTY95} are plotted as filled squares.
}
\end{figure}

To calculate line shapes of the Cs $D$ transitions in liquid helium,
we evaluated Eq. (\ref{defspectrum_1})
by sampling 100 000 helium
configurations, generated according to
the DFT density profiles.
The calculated energy shifts are added to
the observed $D$ lines of free Cs atom
($\lambda=894.9$ nm for $D_1$ and 852.7 nm for $D_2$).
Then, the intensity is estimated by counting number of events
in bins of wavelength $\Delta\lambda=0.1$ nm.
The obtained intensity spectra are shown in Fig.~\ref{fig: No_correlation}.
The $D_1$ line can be well approximated by a single Gaussian,
while the $D_2$ line has a double-peaked structure.
The calculated line shifts and shapes
agree with experimental observations.\cite{KFTY95,KFY96}

\section{Conclusion}
We have developed a simple model to describe atomic spectra of impurities
embedded in the superfluid helium.
Various features in the atomic spectrum of Cs,
including line shifts, broadening, and skewness, are nicely
reproduced in our calculation without any adjustable parameters.
The model is simple enough to apply to more complex chromophores
such as molecules and clusters.
Detailed analysis is found in our recent paper.\cite{NYB02}


\begin{thebibliography}{9}
\bibitem{TV98}
J.~P. Toennies and A.~F. Vilesov,  \Journal{\ARPC}{49}{1}{1998}.
\bibitem{Cal01}
C.~Callegari {\it et al}, Preprint: physics/0109070 (2001).
\bibitem{KFTY95}
T.~Kinoshita {\it et al}, \Journal{\PRA}{52}{2707}{1995}.
\bibitem{KFY96}
T.~Kinoshita {\it et al},  \Journal{\PRB}{54}{6600}{1996}.
\bibitem{Oga99}
S.~Ogata.  \Journal{\JPSJ}{68}{2153}{1999}.
\bibitem{Dup90}
J.~Dupont-Roc {\it et al}, \Journal{J. Low Temp. Phys.}{81}{31}{1990}.
\bibitem{NYB02}
T.~Nakatsukasa, K.~Yabana, and G.~F.~Bertsch, {\em Phys. Rev.} A, in press;
Preprint: physics/0110032.
\end{thebibliography}
\end{document}